\documentclass[aps,prl,superscriptaddress,amsmath,amssymb,floatfix,twocolumn]{revtex4-1}
\usepackage{times}
\usepackage{graphicx}
\usepackage{subfigure}
\usepackage{color}
\usepackage{epstopdf}
\newcommand{\KFS}{\text{K$_{1-x}$Fe$_{2-y}$Se$_2$}}
\newcommand{\AFS}{\text{A$_{1-x}$Fe$_{2-y}$Se$_2$}}

\begin{document}

\title{Orbital-selective Mott Phase in Multiorbital Models for Alkaline Iron Selenides K$_{\rm 1-x}$Fe$_{\rm 2-y}$Se$_{\rm 2}$}

\author{Rong Yu}
\affiliation{Department of Physics \& Astronomy, Rice University, Houston, Texas 77005}
\author{Qimiao Si}
\affiliation{Department of Physics \& Astronomy, Rice University, Houston, Texas 77005}

\begin{abstract}
We study a multiorbital model for the alkaline iron selenides
\KFS\ using a slave-spin method.
With or without ordered vacancies, we identify
a metal-to-Mott-insulator transition at the commensurate filling of six 3d electrons per iron ion.
For Hund's couplings beyond a threshold value,
this occurs via an intermediate orbital-selective Mott phase,
in which the 3d xy orbital is Mott localized while the other 3d orbitals remain itinerant. This phase
is still  stabilized  over a range of carrier dopings.
Our results lead to an overall phase diagram for the alkaline iron selenides,
which provides a unified framework to understand the interplay between the strength of vacancy order
and carrier doping.  In this phase diagram,
the orbital-selective Mott phase provides a natural link between the superconducting
\KFS\ and its Mott-insulating parent compound.
\end{abstract}

\maketitle

{\it Introduction.~} An important question of active discussions is the strength of the electron correlations
in iron-based superconductors~\cite{Kamihara_FeAs,Zhao_Sm1111_CPL08}. The issue is crucial
for understanding the properties
of both the normal and superconducting states of these systems.
For the iron pnictides, the antiferromagnetically ordered metallic ground state
of the parent compounds~\cite{Cruz} may arise either from the Fermi surface nesting of a weak coupling theory~\cite{Dong,Graser},
or from the strong correlation effects associated with the proximity to a Mott transition
and the concomitant quasi-local moments~\cite{Si,Yildirim,Ma,Fang:08,Xu:08,Dai,Uhrig}.
A number of factors argue in favor of the latter picture, including the large spectral weight in the fluctuating magnetic
spectrum~\cite{Liu12}.
For the 11 iron chalcogenides~\cite{Bao11},
both the large ordered magnetic moment and the ordering wave vector
are difficult to understand within the nesting picture.
Also pertinent is the iron oxychalcogenide La$_2$O$_3$Fe$_2$Se$_2$, in which
the Mott insulating behavior has been experimentally identified and theoretically
explained  in terms of the band narrowing effect associated with the expansion of the iron lattice unit
cell~\cite{Zhuetal10}. These results suggest that the incipient Mott picture ~\cite{Si_NJP,Haule09}
is even more pronounced  in the iron chalcogenides.

The recently discovered alkaline iron selenide superconductors~\cite{Guo} \AFS\
(A=K, Rb, Cs, or Tl) shed new light on this issue. In these materials the superconducting $T_c$ is comparable to that
of the pnictides~\cite{Guo,Sun,Krzton-Maziopa,Mizuguchi}, and the superconductivity is near an insulating phase~\cite{Fang,DMWang}.
The insulator is antiferromagnetic with a very large ordered moment~\cite{Bao122,MWang},
and is intimately connected to the ordered iron vacancies~\cite{Fang,Bao122,Ye}.
The lack of hole pockets in the Fermi surface of the superconducting compounds revealed by the ARPES
measurements~\cite{Zhang_Feng,Qian_Ding,Mou} makes the high $T_c$ superconductivity
and the large-moment magnetic order
hardly explainable by the nesting mechanism. Instead, they are more naturally understood in terms of the incipient Mott picture.
For instance, the insulating state is naturally interpreted as a Mott insulator (MI), not only because it would have been metallic
in the absence of interactions but also because the interactions are strong as inferred
from the large ordered moment;
the enhanced interaction effects have been attributed to
the band-narrowing
caused by the ordered iron vacancies~\cite{YuVacOrder11,Zhou,CaoDai}. Various experiments suggest that
the superconducting state is either free of iron vacancies or vacancy disordered, and is intrinsically phase separated
from the vacancy ordered insulating state~\cite{Wang,Li,Chen_Feng}. A key open question is how
the vacancy ordered insulating state connects to superconducting phase.
Elucidating the linkage between the two phases is an important goal
of the present study.

It is well known that tuning either correlation strength or carrier density may induce metal-to-insulator transitions (MITs)
in correlated electron systems~\cite{ImadaRMP}. In multiorbital systems, the physics associated with this transition
is richer than that of one-band systems since both the bandstructure and the correlation strength may be orbital dependent.
An extreme example is the orbital-selective Mott transition (OSMT), for which the Mott transition takes place
at different correlation strengths for different orbitals~\cite{Anisimov,Medici09}. It is believed that the OSMT
occurs in (Ca,Sr)$_2$RuO$_4$.\cite{Anisimov,Medici09,Neupane_Ding}
For iron-based superconductors, strong orbital differences have been suggested in several
systems~\cite{Yu11,Yin_NatMat,Craco}. For the iron pnictides, it was shown theoretically~\cite{Yu12}  that the OSMT
has nearly as competitive a ground-state energy as the other competing phases
but is ultimately not stabilized as a ground state.

In this letter, we investigate the MIT in the
alkaline iron selenides system \KFS\,
using a slave-spin method~\cite{deMedici05,Yu12}. We show that when the Hund's coupling is sufficiently strong,
the Mott localization of the system is always via an OSMP, in which the 3d xy orbital is Mott localized, while the other orbitals
are still itinerant. This OSMP generally exists in both the iron vacancy ordered and disordered cases, and survives
a range of carrier doping. It provides a necessary connection between the vacancy ordered insulating phase and
the metallic normal state above T$_\mathrm{c}$.
Our results allow us to make contact with
recent ARPES measurements in this system~\cite{Yi12}.

{\it Model and method.~}
We consider a multiorbital Hubbard model for the \KFS\
compound. The Hamiltonian reads
\begin{equation}
 \label{Eq:Ham_tot} H=H_0 + H_{\mathrm{int}}.
\end{equation}
Here, $H_0$ contains the tight-binding parameters among the five 3d-orbitals,
\begin{equation}
 \label{Eq:Ham_0} H_0=\frac{1}{2}\sum_{ij\alpha\beta\sigma} t^{\alpha\beta}_{ij} d^\dagger_{i\alpha\sigma} d_{j\beta\sigma} + \sum_{i\alpha\sigma} (\epsilon_\alpha-\mu) d^\dagger_{i\alpha\sigma} d_{i\alpha\sigma},
\end{equation}
where $d^\dagger_{i\alpha\sigma}$ creates an electron in orbital $\alpha=1,...,5$ with spin $\sigma$ at site $i$, $\epsilon_\alpha$
is the on-site energy reflecting the crystal level splitting, and $\mu$ is the chemical potential. We have taken the tight-binding parameters
from Ref.~\onlinecite{Yi12}, and the units are in eV.
$H_{\rm{int}}$
contains on-site Coulomb interactions
\begin{eqnarray}
 \label{Eq:Ham_int} H_{\rm{int}} &=& \frac{U}{2} \sum_{i,\alpha,\sigma}n_{i\alpha\sigma}n_{i\alpha\bar{\sigma}}\nonumber\\
 &+& \sum_{i,\alpha<\beta,\sigma} \left\{ U^\prime n_{i\alpha\sigma} n_{i\beta\bar{\sigma}}\right. 
 + (U^\prime-J) n_{i\alpha\sigma} n_{i\beta\sigma}\nonumber\\
&-&\left.J(d^\dagger_{i\alpha\sigma}d_{i\alpha\bar{\sigma}} d^\dagger_{i\beta\bar{\sigma}}d_{i\beta\sigma}
 -d^\dagger_{i\alpha\sigma}d^\dagger_{i\alpha\bar{\sigma}}
 d_{i\beta\sigma}d_{i\beta\bar{\sigma}}) \right\}.
\end{eqnarray}
where $n_{i\alpha\sigma}=d^\dagger_{i\alpha\sigma} d_{i\alpha\sigma}$. In this model,
$U$, $U^\prime$, and $J$ respectively denote the intraorbital repulsion, interorbital repulsion,
and Hund's rule exchange coulping. We will take $U^\prime=U-2J$.~\cite{Castellani78}

\begin{figure}[t!]
\centering\includegraphics[
width=85mm]{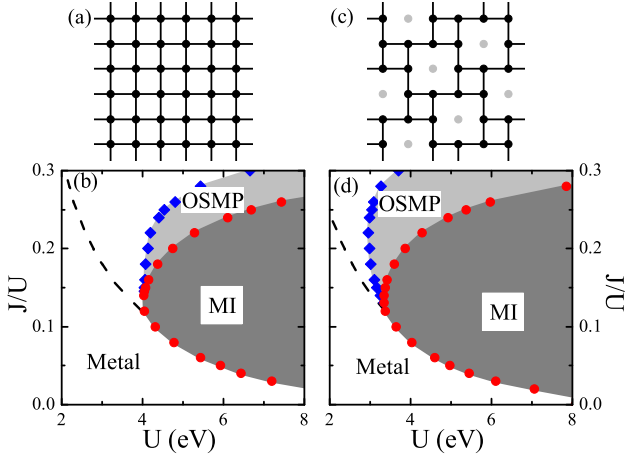}
\caption{(Color online)
a),c): The regular and 1/5-depleted square lattices, respectively corresponding to the alkaline iron selenides with disordered and
the $\sqrt{5}\times\sqrt{5}$ ordered iron vacancies. b),d): Corresponding phase diagrams in the $J$-$U$ plane at commensurate
filling $N=6$ per Fe for the multiorbital model. The dark and light shaded regions respectively refer to the Mott insulator (MI)
and the orbital-selective Mott phase (OSMP). The red circles and blue diamonds respectively denote the Mott transition and
the crossover between the fully itinerant metal phase and OSMP. The black dashed line shows the crossover scale $U^*$ between
the weakly and strongly correlated metals.
}
\label{fig:1}
\end{figure}

The MIT of the above model is studied using a $U(1)$ slave-spin method~\cite{Yu12}.
Here, a slave S=1/2 quantum spin is introduced to carry the charge degree of freedom, and the metallic (Mott insulating) state
corresponds to the magnetically (dis)ordered state of the slave spins with quasiparticle spectral weight in each orbital $Z_\alpha>0$
($Z_\alpha=0$). For simplicity, in the calculation we drop the spin-flip and pair-hopping terms in the interaction $H_{\rm{int}}$.
Including these terms leads to similar results~\cite{Yu11}. We study the MIT on two two-dimensional (2D)
lattices of iron ions: a regular square lattice sketched in Fig.~\ref{fig:1}(a) and a 1/5-depleted square lattice shown in Fig.~\ref{fig:1}(c).
They respectively stand for the completely disordered and the perfect $\sqrt{5}\times\sqrt{5}$ iron vacancies.

\begin{figure}[t!]
\centering\includegraphics[
width=85mm]{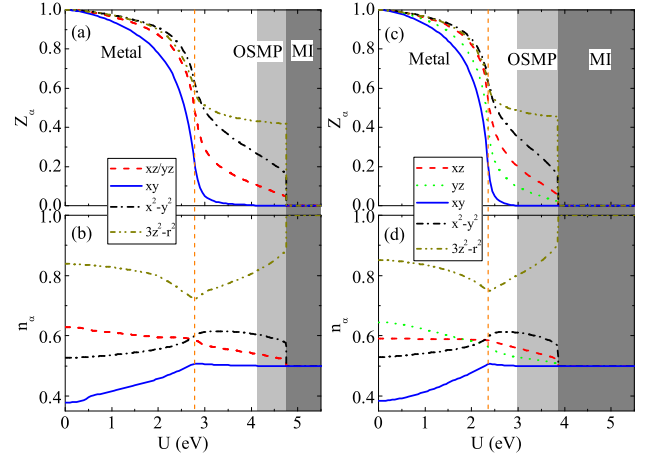}
\caption{(Color online)
(a) and (b): Evolution of orbital resolved quasiparticle spectral weight $Z_\alpha$ (in (a)) and orbital filling factor
(per iron site per spin) with $U$ for the multiorbital model at $N=6$ and $J/U=0.2$ on the regular lattice.
The vertical dashed line indicates the position of $U^*$,
which specifies the dashed lines in  Fig.~\ref{fig:1}(b)(d).
(c) and (d): Same as (a) and (b) but on one of the two inequivalent
sites of a unit cell of the 1/5-depleted lattice.
$Z_\alpha$ for
the xz and yz orbitals switch on the other site, as do the orbital filling factors.
}
\label{fig:2}
\end{figure}

{\it Metal-to-insulator transition.~} The results at the commensurate filling corresponding to $N=6$ 3d electrons per Fe are summarized
in the phase diagrams of Figs.~\ref{fig:1}(b) and (d). In both the vacancy disordered and ordered cases, the system experiences
a Mott transition at $U_{MT}$ from a metal to a MI with increasing $U$. The insulating phase is a low-spin MI for $J/U\lesssim0.01$,
but a S=2 high-spin MI for larger $J$ values. $U_{MT}$ first decreases then increases with increasing $J/U$ ratio.
Such a nonmonotonic behavior is a general feature of systems away from one electron per orbital,
and is also obtained in the five-orbital model for the parent iron pnictides~\cite{Yu12}. When the Hund's coupling is above a threshold ($J/U\gtrsim 0.1$),
the system crosses over from a weakly correlated metal to a strongly correlated metal with increasing $U$.
The onset of this crossover (at $U^*$) is identified by a rapid drop
of $Z_\alpha$ and a kink in
the orbital filling in each orbital, as shown in Figs.~\ref{fig:2}(a)-(d).
In the strongly correlated metallic state, $Z_\alpha$ is strongly orbital dependent. Increasing $U$ does not lead to the simultaneous Mott localization of all orbitals.
The metallic state first crosses over to an intermediate OSMP at $U_{OS}$. The Mott transition then takes place between
the MI and the OSMP at a larger $U$.

Importantly, the phase diagram of the vacancy ordered system is similar to that of its vacancy disordered counterpart.
Quantitatively, $U_{MT}$ and $U_{OS}$ are respectively smaller in the vacancy ordered system, which reflects the
ordered-vacancy-induced reduction in the
kinetic energy and hence enhancement in the correlation effects~\cite{YuVacOrder11}.

{\it Nature of the orbital-selective Mott phase.~} As shown in Figs.~\ref{fig:2}(a) and (c), in the strongly correlated metallic regime, $Z$ in the xy orbital
is suppressed the most, and this orbital is very close to half-filling. Further increasing $U$ results in the Mott localization of the xy orbital
at $U_{OS}$. The other orbitals remain itinerant up to $U_{MT}$. The system is thus in an OSMP for $U_{OS}<U<U_{MT}$. We now turn
to discussing the factors that stabilize the OSMP. For simplicity, we limit our discussion to the vacancy disordered case.
The vacancy ordered case is qualitatively similar.

\begin{figure}[t!]
\centering\includegraphics[
width=85mm]{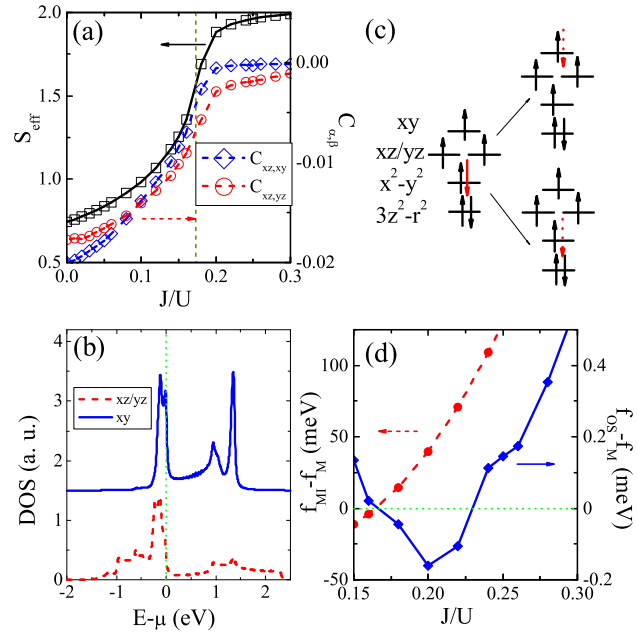}
\caption{(Color online)
(a): Evolution of the effective moment $S_\mathrm{eff}$ and interorbital correlations $C_{xz,xy}$ and $C_{xz,yz}$
with $J/U$ at $N=6$ and $U=3$ eV on the regular lattice.
The vertical dashed line indicates the position of $U^*(J)$.
(b): Orbital projected density of states in xy and xz/yz orbitals
at the noninteracting limit ($J$=$U$=0) of the same model.
The curve for the xy orbital is shifted upward for clarity.
(c): Sketch of the two hopping processes of a charge excitation on the high-spin ground-state configuration that avoid penalty
from the repulsive interactions. (d): The differences in free energy density as a function of $J/U$ at $U=4.2$ eV, showing the competition
among the metallic (M), Mott insulating (MI), and orbital-selective Mott (OS) solutions.
}
\label{fig:3}
\end{figure}

We start from the physics that governs the crossover between the weakly and strongly correlated metals. Fig.~\ref{fig:3}(a)
plots the effective magnetic moment $S_\mathrm{eff}=\sqrt{\langle {(S^z)}^2\rangle}$ as a function of $J/U$. It rapidly increases
when the system passes through the crossover.
In the strongly correlated metallic phase, $S_\mathrm{eff}\approx2$, indicating that the $S=2$ high-spin configuration, promoted
by the Hund's coupling,
is dominant in this regime.

The Hund's coupling also suppresses the inter-orbital correlations
$C_{\alpha,\beta}=\langle n_{\alpha} n_{\beta}\rangle - \langle n_{\alpha}\rangle \langle n_{\beta}\rangle$, as shown in Fig.~\ref{fig:3}(a).
Together with the crystal level splitting, this effectively decouples the xy orbital from the others because in
\KFS\
the xy orbital is the topmost level and is well separated from the others. For the same reason, it is easier to stabilize the xy orbital
to be at half-filling for an overall filling of six electrons per Fe, as shown in Fig.~\ref{fig:2}(b)
[and Fig.~\ref{fig:2}(d) for the vacancy-ordered case]. Compared to the degenerate xz/yz orbitals, which are also close to half-filling, the threshold value for the Mott localization in the non-degenerate xy orbital is smaller.
Moreover, in the noninteracting limit the density of states (DOS)
projected to the xy orbital is narrower than those of other orbitals. For instance, as shown in Fig.~\ref{fig:3}(b), the ratio of the
width of DOS for the xy orbital to that for the xz/yz orbitals is about 0.6. (This is also the case for the vacancy ordered
model; see the figure in the Supplementary Material.) This ratio is smaller than that for
the LaOFeAs system, which is about 0.7.
Hence
in
\KFS\,
the xy orbital contains less kinetic energy.
Taking into account all the three factors, we find that in the strongly correlated metallic state, it is much easier to drive the xy orbital
toward the Mott localization. The result is the OSMP.

The threshold interaction for the OSMP, $U_{OS}$, shows a strong dependence on $J$ especially when $J/U$ is large.
This seems counterintuitive; one could expect that, if the xy orbital is fully decoupled from the others, $U_{OS}$ should approach
the critical
$U$ of a single-band Hubbard model, and hence should not depend on $J$. To understand the behavior of $U_{OS}$,
we examine the propagation of a charge excitation by assuming both $U$ and $J$ are large so that we may
take the ground-state configuration to be the $S=2$ high-spin state. A charge excitation with one more electron
filled in the ground state can propagate via hopping to neighboring sites.
$K_{xy}$ ($K_{\overline{xy}}$) denotes the kinetic energy gain associated with the hopping processes (not) involving the xy orbital. Two representative hopping processes that do not disturbe the high-spin ground-state configuration are illustrated in Fig.~\ref{fig:3}(c).
Note that $K_{xy} < K_{\overline{xy}}$, not only because the xy orbital has a narrower (non-interacting) bandwidth
but also because the Hund's coupling suppresses the interorbital fluctuations.
The Mott gap associated with either process is estimated as
$\Delta_a = E(N+1)+E(N-1)-2E(N) \approx U-3J-K_a$,
where $a={xy,\overline{xy}}$. The difference in the kinetic energy gains leads to two different Mott gaps.
$U_{OS}$ ($U_{MT}$) can be estimated as the $U$ value where the Mott gap $\Delta_a$ vanishes.
Hence we have $U_{OS} \sim K_{xy}/(1-3J/U)$ and $U_{MT} \sim K_{\overline{xy}}/(1-3J/U)$, both of which increase with $J/U$.
This general consideration is consistent with our calculated phase boundaries for sufficiently large $J/U$. Interestingly, in this regime,
increasing $J$ at a fixed $U$ leads to a delocalization from MI to OSMP, and then to the metallic phase. This is further confirmed
by comparing the free energies of the three states, as shown in Fig.~\ref{fig:3}(d). Note that in the metallic phase, the inter-orbital
correlations between the xy orbital and others are substantially suppressed, but remain nonzero.
The above argument on the behavior of $U_{OS}$ does not hold for smaller
$J/U$ where $U_{OS}$ is close to $U^*$. In this regime, since the ground
state mixes both high- and low-spin configurations, several mechanisms
favoring either increasing or decreasing $U_{OS}$ compete. As a result,
$U_{OS}$ shows complicated, even nonmonotonic, $J$ dependence (Fig.~\ref{fig:1}(b)).

\begin{figure}[t!]
\centering\includegraphics[
width=80mm]{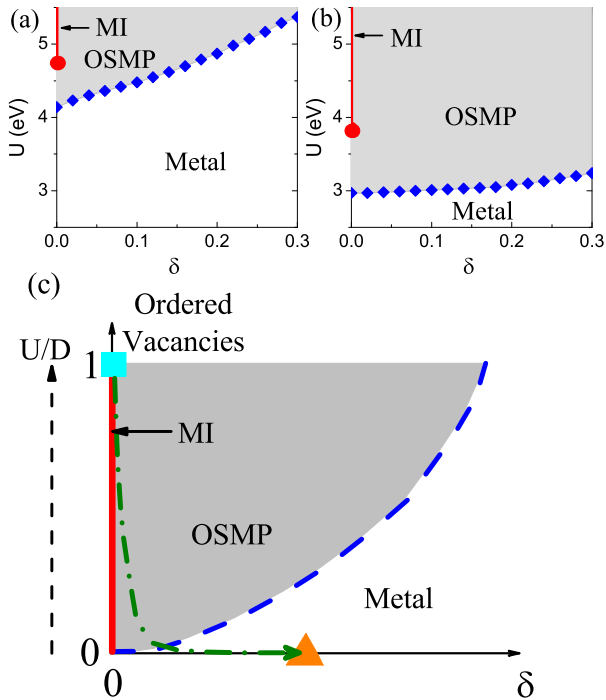}
\caption{(Color online)
(a) and (b): Phase diagrams with $U$ and carrier doping concentration $\delta$ for the multiorbital model at $J/U=0.2$ on the regular
(in (a)) and the 1/5-depleted (in (b)) lattices, respectively. In either diagram, the large red dot refers to the Mott transition,
and the blue diamonds shows the orbital-selective Mott transition. (c): Sketch of a material-based phase diagram
 in the plane of carrier doping $\delta$ and ordered vacancies. The vacancy order parameter has been
scaled to be between 0 and 1 (see text).
The vacancy ordered insulating state is located as the cyan square in this phase diagram.
The superconducting state is tentatively placed as the orange triangle on the $\delta$ axis.
The dash-dot line shows a possible route to connect the two phases.
For realistic parameters, the Mott transition point is close to the origin, but could be either above or below it.
}
\label{fig:4}
\end{figure}

{\it Orbital-selective Mott phase at finite dopings.~}
Unlike the MI, which exists only at a commensurate filling, the OSMP can be stabilized at incommensurate fillings if the chemical potential
of the itinerant carriers falls inside the Mott gap of the localized orbital. In Figs.~\ref{fig:4}(a) and (b) we show the $U$ vs.
carrier doping concentration ($\delta=N-6$) phase diagrams for both the vacancy disordered and ordered cases.
In both systems, the Mott transition takes place between a MI and an OSMP at the commensurate filling $\delta=0$,
and an OSMT between the OSMP and the metal extends to nonzero doping concentrations. In the vacancy ordered case,
both the Mott transition and the OSMT take place at lower $U$ values compared to the vacancy disordered case,
reflecting the enhanced correlations due to ordered vacancies.

{\it Unified phase diagram for alkaline iron selenides.~}
In light of the above two phase diagrams, we propose a material-based unified phase diagram for both the vacancy
ordered and disordered compounds. In this diagram, sketched in Fig.~\ref{fig:4}(c), the horizontal axis refers
to the carrier doping $\delta$, and the vertical axis stands for the strength of the vacancy order. In general, an
\KFS\ system
contains both vacancy ordered and disordered regimes. We may parameterize the strength of the vacancy order from 0 to 1,
according to the volume fraction of the vacancy ordered regime (or, alternatively, to the potential strength
of a virtual Fe atom,
with the vacancy corresponding to an infinite potential).
The two limiting cases
along the vertical axis in the phase diagram, 0 and 1, are obtained from Figs.~\ref{fig:4}(a) and (b) at a fixed $U$ (which takes the value
in real materials), respectively.
The remaining part of the phase diagram can then be constructed by interpolating between the results in Figs.~\ref{fig:4}(a)
and (b) at the same $U$.
The resulting diagram consists of a MI, an OSMP, and a metallic phase. The profile is similar to what are shown in
Figs.~\ref{fig:4}(a) and (b) because, effectively, the
ordered vacancies enhance the correlation $U/D$. (Here $D$ is a characteristic bandwidth of the multiorbital system.)
The insulating compound with the $\sqrt{5}\times\sqrt{5}$ vacancy order is located at $\delta=0$ and vacancy order 1 (square cyan symbol in Fig.~\ref{fig:4}(c)).
On the other hand, we tentatively place the superconducting phase of the superconducting compounds at ambient pressure to be
on the $\delta$ axis in the metallic state close to the OSMT (triangular orange symbol in Fig.~\ref{fig:4}(c)).
We see that one physical trajectory going from
the insulating phase
to the superconducting one is for the dopants to both introduce extra carriers and suppress the vacancy order.
During the evolution of the system,
the OSMP is an unavoidable intermediate phase connecting the insulating and superconducting states.
A recent ARPES measurement ~\cite{Yi12}
has provided evidence that the superconducting phase of \KFS\
is indeed close to the OSMT.

In summary, we have studied the metal-to-insulator transition in the five-orbital Hubbard model for the alkaline iron selenides
\KFS\  with and without ordered vacancies.
We find that the Mott localization of the system is via an intermediate orbital-selective Mott phase, in which the 3d xy orbital
is localized while the other 3d orbitals are itinerant. The orbital-selective phase is stabilized by the combined effect of the orbital dependence
in the bandwidths and the pinning of the xy orbital to half-filling due to both the Hund's coupling and crystal level splittings.
It persists over a range of carrier dopings. Finally, we have proposed a unified phase diagram for the alkaline
iron selenides, with the orbital-selective Mott phase providing the link between the insulating and superconducting compounds.

We thank Z.\ K. Liu, D.\  H.\ Lu, Z.\ X.\ Shen, L.\ L.\ Sun, and M.\ Yi for useful discussions.
This work has been supported by the NSF Grant No. DMR-1006985 and the Robert A.\ Welch Foundation
Grant No.\ C-1411. Q. S. acknowledges the hospitality of the Aspen Center for Physics (NSF Grant
No. 1066293) and the Institute of Physics of Chinese Academy of Sciences.
This work was reported in Ref.~\cite{Yu_APS12}.
Complementary theoretical results on different systems using different methods have subsequently been reported in
Refs.~\cite{Yin12,Bascones12}.

\newpage

\onecolumngrid

\section*{{\Large Supplementary Material}}

\begin{figure}[h!]
\centering\includegraphics[
width=85mm]{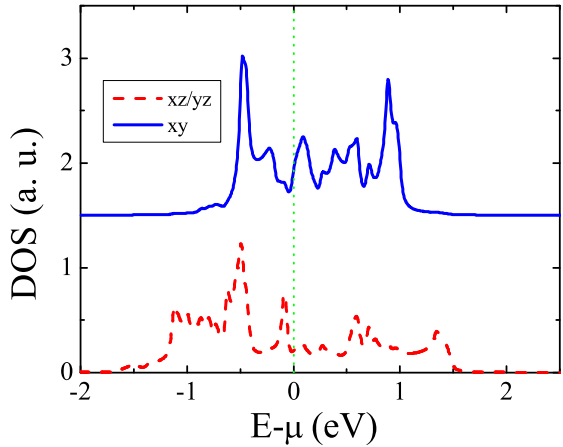}
\caption{(Color online)
Orbital projected density of states (DOS) in xy and xz/yz orbitals
at the noninteracting limit ($J$=$U$=0) of the vacancy ordered model at $N=6$. The curve for the xy orbital is shifted upward for clarity.
The DOS has been averaged over the two inequivalent sites of a unit cell.
}
\label{fig:S1}
\end{figure}

Fig.~\ref{fig:S1} shows the orbital projected density of states (DOS) in xy and xz/yz orbitals at the noninteracting limit ($J$=$U$=0) of the vacancy ordered model. The DOS has been averaged over the two inequivalent sites of a unit cell. In the vacancy ordered model, the DOS in the xy orbital is narrower than those in the other Fe 3d orbitals, and the ratio of the width of DOS for the xy orbital to that for the xz/yz orbitals is about 0.6. These are both similar to the vacancy disordered model.

\end{document}